\documentclass[12pt]{article}
\usepackage[utf8]{inputenc}
\usepackage[english]{babel}
\usepackage{amsmath,amsfonts,amssymb,bm,CJK}
\usepackage{graphics,epstopdf,ifpdf}
\usepackage{mathrsfs,textcomp,multicol}
\usepackage{indentfirst,fancyhdr,upgreek}
\usepackage[columnwise]{lineno}
\usepackage[bookmarksnumbered=true,bookmarksopen=true,colorlinks=true,pdfborder={0 0 1},
            urlcolor=blue,linkcolor=blue,anchorcolor=blue,citecolor=blue]{hyperref}

\usepackage{indentfirst}
\usepackage{graphicx}
\usepackage{dcolumn}
\usepackage{bm}
\usepackage{lipsum}
\usepackage{slashed}
\usepackage{enumitem}
\usepackage{txfonts}
\usepackage{esint}
\usepackage{xcolor}
\usepackage{mathtools}
\usepackage{transparent}
\usepackage{upgreek}
\usepackage{tabularx}
\usepackage{csquotes}
\usepackage{geometry}
\usepackage[font=footnotesize,labelfont=bf,labelsep=period]{caption}
\usepackage{tikz}
\usepackage[compat=1.1.0]{tikz-feynman}
\tikzfeynmanset{/tikzfeynman/warn luatex=false}
\usetikzlibrary{arrows,automata}
\usetikzlibrary{matrix}
\addto\captionsenglish{}

\usepackage[backend=biber, style=phys,articletitle=false,biblabel=brackets,
            chaptertitle=false,pageranges=false]{biblatex}      
\usepackage{definitions}

\begin{document}
\title{ {\tt SANC} Monte Carlo programs for small-angle Bhabha scattering\footnote{Supported by the Russian Science Foundation (project No. 22-12-00021)}
}

\author{ 
	\parbox{\linewidth}{\centering
  A.B.\,Arbuzov$^{1,3}$, S.G.\,Bondarenko$^{1,3}$, I.R.\,Boyko$^{2}$,
  L.V.\,Kalinovskaya$^{2}$, A.A.\,Kampf$^{2}$,\\
  R.R.\,Sadykov$^{2}$,
  V.L.\,Yermolchyk$^{2,3}$\thanks{E-mail:~Vitaly.Yermolchyk@jinr.ru}.
}
}

\date{}

\maketitle

\vskip 0.3cm

\begin{center}
{
$^1$ \it Bogoliubov Laboratory of Theoretical Physics, JINR, Dubna, 141980 Russia\\
$^2$ \it Dzhelepov Laboratory of Nuclear Problems, JINR, Dubna, 141980 Russia \\
$^3$ \it Dubna State University, Dubna, 141982 Russia
 }
\end{center}

\begin{abstract}
Luminosity monitoring at e+e- colliders is investigated using
 {\tt SANC} Monte Carlo event generator 
 {\tt {ReneSANCe}} and integrator 
 {\tt {MCSANC}} for simulation 
of Bhabha scattering at low angles.  
Results are presented for center-of-mass energies of the $Z$ boson resonance and 240~GeV for the conditions 
of typical luminosity detectors.
It is shown that taking into account bremsstrahlung events with extremely low electron scattering angles 
is relevant to match the precision tags of the  future electron-positron colliders. 
\end{abstract}


\section{Introduction}

Luminosity monitoring is the standard task for all collider experiments.
One of the traditional processes for high-precision luminosity measurements 
at electron-positron colliders is Small-Angle Bhabha Scattering (SABS).
This process has a clean detector signature and 
very large cross section which sharply increases at small scattering angles.
From the theoretical point of view
it is almost a pure QED process and thus can be described very accurately
within perturbative quantum field theory.
SABS occupies a special place in the physics programme of future $e^+e^-$ colliders
like  FCCee~\cite{FCC:2018evy} and CEPC~\cite{CEPCStudyGroup:2018ghi}.
Given the extremely large expected statistics, the luminosity measurement 
with precision $10^{-4}$ or better is necessary.
The theoretical accuracy for SABS calculations must be significantly 
better than this target precision in order not to spoil the resulting
uncertainty.

The most advanced codes  
for theoretical estimation of luminosity with the help of {\tt SABS}  are
{{\tt{BabaYaga}}} \cite{CarloniCalame:2003yt,BALOSSINI2006227,BALOSSINI2008209,CARLONICALAME200116,CarloniCalame:2000pz}, 
{{\tt{BHLUMI}}} \cite{Jadach:1996is}.
The  Monte Carlo (MC) generator {\tt{BHLUMI}} is a pure QED tool
and its theoretical uncertainty is estimated to be about 
 0.037$\%$, see Table 2 in~\cite{Jadach:2021ayv}.
In that paper the future prospects of theoretical precision
 $1 \times 10^{-4}$  was presented
for luminosity measurement 
at the future colliders at
the $Z$ peak.

The new release of {\tt {BabaYaga}} 
\cite {CarloniCalame:2019nra} is accounting for the various sources of radiative corrections, i.e. QED, (electro)weak and higher order effects.
This generator is mainly intended for large angle Bhabha scattering, with
theoretical errors of about  0.1$\%$.

\begin{figure}[!h]
\includegraphics[clip,width=\linewidth,trim={0cm 5cm 0cm 0cm}]
{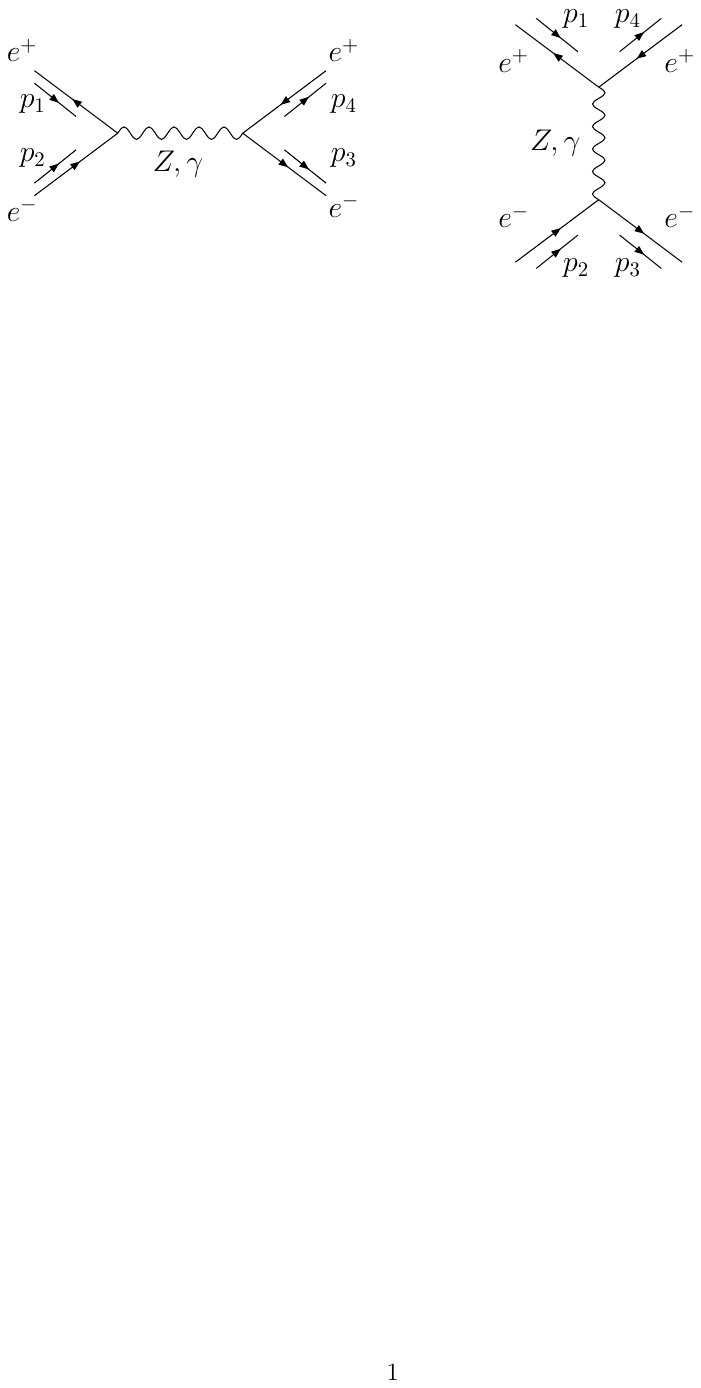}
\vspace*{-13cm}

\caption{The $s$ and $t$ channels of Bhabha processes at lowest order.}
    \label{diagrams}
\end{figure}

In this paper we present a study of SABS based on {\tt MCSANC} integrator ~\cite{Bondarenko:2013nu}
and {\tt ReneSANCe} generator \cite{Sadykov:2020any}.
The process of 
polarized Bhabha scattering (see Fig.~\ref{diagrams})
\begin{eqnarray}
\label{bhabha} \nonumber
e^{+}(p_1) + e^{-}(p_2) \rightarrow 
e^-(p_3) + e^+(p_4) + (\gamma (p_5))
\end{eqnarray}
was calculated at the complete one-loop electroweak level \cite{Bardin:2017mdd}. 
In addition here we consider the higher order corrections by $\delta\rho$ parameter, which are necessary to meet the high-precision requirements of the future $e^+e^-$ experiments.
The details of Bhabha scattering implementation into MC {\tt ReneSANCe} are described in \cite{Sadykov:2020any}.  

The aim of the present paper is to report on the study of 
the  Bhabha scattering  cross section at arbitrarily small or even vanishingly small scattering angles. 
The contribution from electron scattering at very small angles introduces additional, potentially sizeable, effect in the theoretical interpretation of the measured  {\tt SABS} cross section value.
We provide the advanced assessment 
of SABS events with scattering angles under 10 mrad.
Earlier, this kinematic region could be described by {\tt BHAGEN-1PH} \cite{Caffo:1996mi}, however the calculations were limited to the contribution of hard photon Bremsstrahlung.

The outline of the paper is as follows. 
In Section \ref{Section2} we show 
the comparisons with the results of alternative 
MC code 
in the conditions and setup 
of the CERN Workshop
~\cite{Jadach:1996gu}.
In Section \ref{Section3} we give numerical results for the integral cross sections and angular event distributions of    experimental interest in {\tt SABS}. We also discuss different  sources of radiative
corrections and study the effect from 
the minimum cut-off on electron scattering angle.

\section{Cross-check with the 1996 LEP Workshop
\label{Section2}}

To verify 
the technical precision of our codes
we produced the tuned comparison
with results presented in the proceedings of the CERN Workshop~\cite{Jadach:1996gu} devoted to event generators for Bhabha scattering at LEP
for the 
non-calorimetric event selection called BARE1 and the calorimetric one called CALO1.
All numbers are produced within the setup of this workshop
for the ${\cal{O}} (\alpha)$ matrix element 
without contribution of the  $Z$  exchange, up-down interference and vacuum polarization
and 
with various  values of the energy-cut $z_{\rm min} = s'/s$, where $s'$ is the collision energy 
after initial state radiation (ISR). 
Table \ref{renesancevsbhlumiBIG}  shows a good
agreement
within numerical precision.  

\begin{table*}[!h]
	\caption{Comparison of 
 BARE1 and CALO1 for the ${\cal{O}}
 (\alpha)$ matrix element. $Z$  exchange, up-down interference and vacuum polarization 
 are switched off. The center of mass energy is 
 $\sqrt{s} = 92.3$ GeV. The results are shown with various  values of the energy-cut $z_{\rm min} = s^{'}/s$. }
  \centering\small
 	\begin{tabular}{|l||l|l|l|l|}
		\hline
		$z_{\rm min}$ & {\tt ReneSANCe}  & {\tt BHLUMI} & {\tt {\tt ReneSANCe}} & {\tt BHLUMI}
		\\ \hline
		&\multicolumn{2}{c|}{BARE1: $\sigma$ [nb] } & \multicolumn{2}{c|}{CALO1: $\sigma$ [nb] }\\
		\hline
		$.100$ & 166.01(1) & 166.05(2) & 166.33(1) & 166.33(2)\\
		\hline
		$.300$ & 164.71(1) & 164.74(2) & 166.05(1) & 166.05(2)\\
		\hline
		$.500$ & 162.19(1) & 162.24(2) & 165.26(1) & 165.29(2)\\
		\hline
		$.700$ & 155.41(1) & 155.43(2) & 161.77(1) & 161.79(2)\\
		\hline
		$.900$ & 134.36(2) & 134.39(2) & 149.91(1) & 149.93(2)\\
		\hline
	\end{tabular}

	\label{renesancevsbhlumiBIG}
\end{table*}

\section{Numerical results \label{Section3}}

All results were obtained in the $\alpha(0)$ electroweak scheme using the   set of input parameters listed in  Table~\ref{parameters}. 
\begin{table}[ht!]
    \caption{Input parameters.}
    \centering
    \begin{tabular}{lcllclc}
    $\alpha^{-1}(0)$ &=& 137.035999084    &            & &                & \\
    $M_W$            &=& 80.379 GeV       & $\Gamma_W$ &=& 2.0836 GeV     & \\
    $M_Z$            &=& 91.1876 GeV      & $\Gamma_Z$ &=& 2.4952 GeV     & \\
    $M_H$            &=& 125.0 GeV        & $m_e$      &=& 0.51099895 MeV & \\
    $m_\mu$          &=& 0.1056583745 GeV & $m_\tau$   &=& 1.77686 GeV    & \\
    $m_d$            &=& 0.083 GeV        & $m_s$      &=& 0.215 GeV      & \\
    $m_b$            &=& 4.7 GeV          & $m_u$      &=& 0.062 GeV      & \\
    $m_c$            &=& 1.5 GeV          & $m_t$      &=& 172.76 GeV.    &
    \end{tabular}
    \label{parameters}
\end{table}

In addition, the following conditions were taken into account:
\begin{itemize}
    \item electrons were allowed to scatter by any angle, down to zero,
    \item luminosity acceptance was assumed $ 30 ~\mbox{mrad} < \theta < 174.5~\mbox{mrad}.$
\end{itemize}

To demonstrate {\tt ReneSANCe} capabilities, we generated 100 million events for the Bhabha cross section for the two c.m.s. energies $\sqrt{s}=$ 91.18 GeV and 240 GeV,
where each arm of the luminometer registered an energy shower from an electron or photon.
We do not apply any restrictions on the minimum scattering angle of an electron,
i.e. the electron can scatter down to zero.  

We used two different setup for event selection (ES) called  {\bf ES-BARE} (non-calorimetric) and {\bf ES-CALO} (calorimetric).
In the {\bf ES-BARE} case we define the SABS cross section by choosing events where each arm of the calorimeter is hit by an electron or positron.
These electrons or positrons 
must have an energy of at least half the energy of the beam ($E_{\rm beam}$). 
For the {\bf ES-CALO} setup we consider a calorimetric detector that can't distinguish  electrons from photons. 
 In other words, the cross section is determined by events in which each arm of the calorimeter is hit by either a photon or an electron carring at least  half of the beam energy.

\subsection{Different radiative correction contributions}

To present the main sources of theoretical uncertainties of the one-loop cross section $\sigma^{\mathrm{1-loop}}$, we divide it into gauge-invariant subsets.
When evaluating the contribution cross section at the Born level (leading order, LO) $\sigma^{\rm Born}$, both photon and Z-boson exchanges are taken into account.
In order to quantify the impact of different contributions,
we divide them into  gauge-invariant subsets: QED one-loop corrections $\sigma^{\mathrm{QED}}$, 
the vacuum polarization 
contribution $\sigma^{\rm VP}$, and
  the pure weak contribution $\sigma^{\rm weak}$  as the difference between the complete one-loop electroweak
 correction and the pure QED part $\sigma^{\mathrm{QED}}$ of it.
The leading higher-order (ho) corrections
we denoted as  $\sigma^{\rm ho}$.

In Table \ref{ResultsES-BARE} we show the results of the various radiative contributions to the total cross section for the Z-pole and $\sqrt{s}=240$ GeV
and evaluate corresponding relative corrections as
$\delta = {\sigma^{\text{contr.}}}/{\sigma^{\text{Born}}}$.

\begin{table}[h!]
        \caption{The results of the various radiative contributions to the total cross section for the Z-pole and $\sqrt{s}$ = 240 GeV for {\bf ES-BARE}.}
	\centering
		\begin{tabular}{lcc}
			\hline\hline
			$\sqrt{s}$, GeV & $91.18$ & $240$ \\
			\hline
   
			$\sigma^{\rm Born}$, pb & $135008.970(1)$ & $19473.550(1)$ \\

			$\delta^{\rm one-loop}$, \% & $-1.562(1)$ & $-0.821(1)$\\
      		$\delta^{\rm total}$, \% & $-1.420(1)$ & $-0.574(1)$\\
                \hline
			$\delta^{\rm QED}$, \% &  $-6.296(1)$ & $-7.002(1)$\\
			$\delta^{\rm VP}$, \% & $4.6527(1)$ & $6.1866(1)$ \\
			$\delta^{\rm weak}$, \% & $0.0088(1)$ & $-0.0064(1)$\\
			$\delta^{\rm ho}$, \%&  $0.1418(1)$  & $0.2475(1)$  \\
			\hline\hline
		\end{tabular}
        \label{ResultsES-BARE}
\end{table}

The leading higher-order EW corrections $\delta^{\rm ho}$ 
to SABS are included in our calculations through the $\Delta\alpha$ and $\Delta\rho$ parameters.
 A detailed description of our implementation of this contribution was 
 presented in~\cite{Arbuzov:2021oxs}.  
 At two-loop level the above corrections consist of the EW at ${\cal{O}}(G_{\mu}^2)$ and the mixed EW$\otimes$QCD at ${\cal{O}}(G_{\mu}\alpha_s)$ parts.
For SABS the bulk of the considered higher-order
effects is due to running $\alpha$.

\subsection{{\tt SABS}, analysis of events for {\bf ES-BARE} and {\bf ES-CALO} setups}

Another possible bias to the luminosity measurement arises from events where an electron is scattered at a very small angle and escape detection. Such events can be accepted by a luminometer due to energetic photons radiated at angles large enough to be detected in the detector. This effect would lead to a bias in luminosity measurements if the data is analyzed with an MC tool which uses a minimum scattering angle cut-off.

We use {\tt MCSANC} integrator to compare results from {\bf ES-BARE} setup (ignoring photons) with {\bf ES-CALO} setup in which calorimeter can be hit by either an electron or a photon.
  The presence of a high-energy photon provides a natural regularisation of divergence at zero electron scattering angles.
  Althoug electrons are allowed to be scattered  by zero angle, the number of such events is small because
  of the requirement to have an energetic photon within the acceptance of the  calorimeter.
  We define the luminosity acceptance in the range of 30 mrad to 10 degrees (174.5 mrad),
which is typical for LEP detectors as well as for future $e^+e^-$ colliders like FCCee, CEPC and ILC.

This effect is presented in Table \ref{ResultsBARECALO}.
We observe that the  {\bf ES-CALO} cross-section at $\sqrt{s}=$ 91.18 and 240 GeV is $3\%$ larger than {\bf ES-BARE} Bhabha cross-section, when both beam particles must hit the luminometer.
The largest part of the difference is 
due to the events with collinear photon 
or due to the events in which electron is scattered by an angle
larger than the luminosity acceptance, while hard ISR photon hits the luminometer.
Such effect can not introduce any experimental bias because the electron can be detected by large-angle calorimeters and the process can be simulated by any Bhabha generator.

Additionally, it was  found that approximately $1.4$ permille of the total cross section for both energies is represented  by events with electron scattering angles below the given luminometer acceptance angle 30 mrad. 
The size of this effect $\Delta^{\rm QED}({\rm \vartheta < 0.030})$ can be derived from Table \ref{ResultsBARECALO} as the difference between 
$\delta_3^{\rm QED}$ and $\delta_2^{\rm QED}$.
Note that only the technical uncertainty of numerical integration are shown
in the Table, estimates of the corresponding theoretical uncertainties will
be presented elsewhere.
\begin{table}[h!]
 \caption{Born cross sections and relative corrections  for $\sqrt{s}$ = 91.18 GeV and 240 GeV.
 Here     $\delta_1^{\rm QED} = \delta$({ES-BARE}) is the QED correction
 for {\bf ES-BARE} setup,~
      $\delta_2^{\rm QED} = \delta$({ES-CALO},\ $\vartheta > 0.030)$ is the QED correction
     for the {\bf ES-CALO} setup with electron scattering angles larger than the minimum 
     luminosity acceptance, and $\delta_3^{\rm QED} = \delta$({ES-CALO}) is the QED correction
 for {\bf ES-CALO} setup with arbitrary electron scattering angles. 
 }
\begin{center}
 \begin{tabular}{lcc}
	\hline\hline
	$\sqrt{s}$, GeV & $91.18$ & $240$\\ 
    \hline
    $\sigma^{\rm Born}$, pb             & $135008.970(1)$ & $19473.550(1)$ \\
    $\delta_1^{\rm QED}
    $, \% &  $-6.296(1)$    & $-7.002(1)$     \\
        $\delta_2^{\rm QED}
    $, \% &  $-3.618(1)$    & $-3.986(1)$      \\ 
     $\delta_3^{\rm QED}   
   $, \% &  $-3.488(1)$           & $-3.854(1)$\\
   \hline
   $\Delta^{\rm QED}({\rm \vartheta < 0.030})$   & $1.30(1)\times 10^{-3}$& $1.32(1)\times 10^{-3}$\\
   \hline \hline
	\end{tabular}
\end{center}

   \label{ResultsBARECALO} 
\end{table}

\subsection{Angular distributions} 

In the following, we illustrate the numerical results by the example of several angular distributions obtained with the MC 
generator {\tt ReneSANCe}.
  We consider the distribution of electron scattering angles between
  the outgoing electron and the incoming electron
  as well as the distribution by the angle at which the photon was emitted.

We present angular distributions of two types:
\begin{itemize}
\item[a)]
distribution 
of events by scattering angle
of the Bremsstrahlung photon 
$\vartheta_{15}=\vartheta_\gamma$, i.e., the angle between  particle $p_1$ (initial positron)
and particle $p_5$ (photon),
\item[b)]
distribution 
of events by positron scattering angle
$\vartheta_{14}$, i.e., the angle between  particle $p_1$
and particle $p_4$.
\end{itemize}

\underline{$Z$ resonance} \\

\begin{figure}[h!]
    \begin{tikzpicture}
        \matrix[matrix of nodes]{
        \includegraphics[scale=0.8]{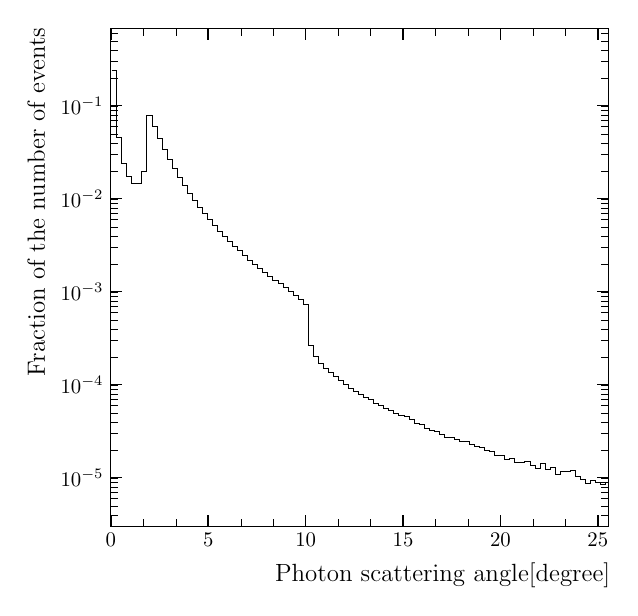} & \includegraphics[scale=0.8]{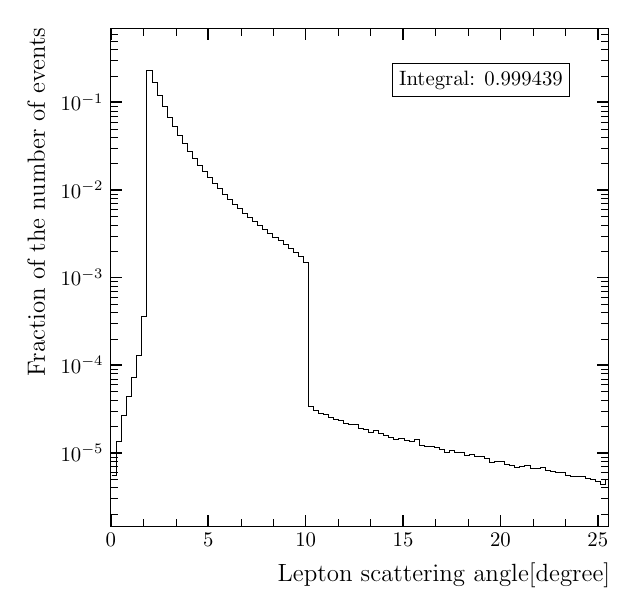}\\}; 
    \end{tikzpicture}
    \caption{Angular distributions of type a) on the left side and type b) on the right side for $\sqrt{s} = 91.18$ GeV.}
    \label{fig:1}
\end{figure}

Figure ~\ref{fig:1} presents the angular distributions of type a) on the left side and of type b) on the right side for c.m.s. $\sqrt{s} = 91.18$ GeV.
 The vertical axes show the relative fraction of events in the given bin.
  The sum of all events is normalized to 1.0 and 
  the numbers in the frames  
  show the fractions of events within the range of a given plot.
As can be seen from the plots,  the event yield  vanishes
  when lepton scattering angle  approaches zero. 
Sharp edges at 1.7 and 10 degrees correspond to acceptance of luminometer. Events with leptons scattered beyond the luminometer acceptance correspond to detection of energetic photons.

\begin{figure}[ht!]
    \begin{tikzpicture}
    \matrix[matrix of nodes]{
    \includegraphics[scale=0.8]{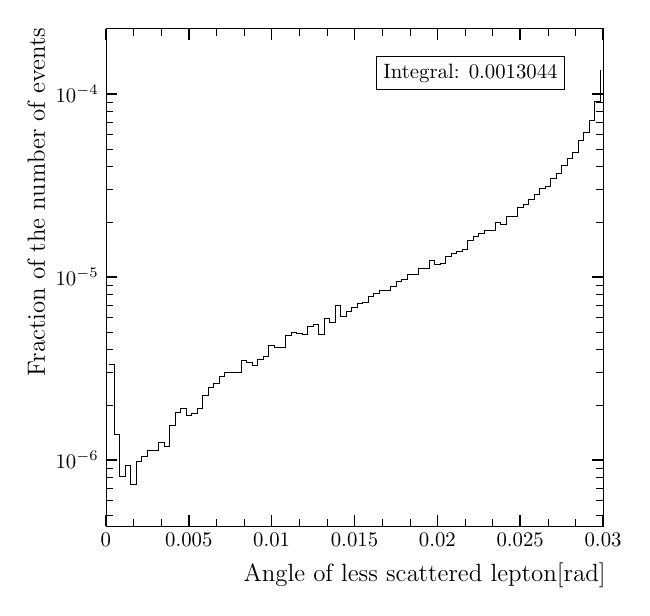} &
    \includegraphics[scale=0.8]{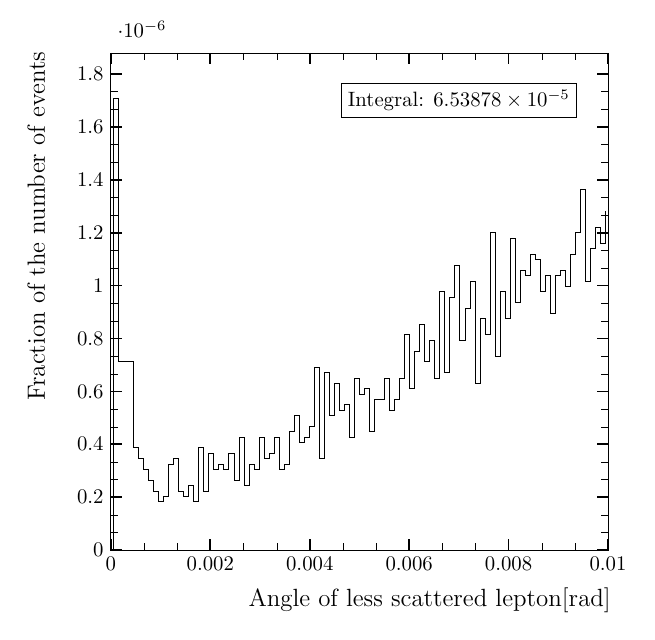}&\\};
    \end{tikzpicture}
    \caption{Angular distributions of type b) in a 
    range $(0, 30)$ mrad (left) and $(0, 10)$ mrad (right) for $\sqrt{s} = 91.18$ GeV.}
    \label{fig:11}
\end{figure}

In Figure 
\ref{fig:11}
  we show the distributions of electrons scattered at the angle less than the acceptance of the luminosity calorimeter. 
  The peak at nearly zero electron scattering angles is due to the terms proportional to $m_e^2/t^2$ (here $t$ is the square of the electron momentum transferred), which are present in the differential cross section of the radiative Bhabha process, see, e.g., ref.~\cite{Arbuzov:1997pj}.
The total fraction of the events within this 
angular range $(0,30)$ mrad is about 1.3 permille. 
For the angular range 0-10 mrad the relative event yield is  $0.65\cdot 10^{-4}$.
 Therefore the MC generator cut-off on electron (and positron) scattering angles somewhat less than 10 mrad 
 would be safe if the experimental systematic error on luminosity measurement is expected at the level of $10^{-4}$.
 \\
  
\underline{Center of mass energy 240 GeV}\\

\begin{figure}[h!]
    \begin{tikzpicture}
        \matrix[matrix of nodes]{
        \includegraphics[scale=0.8]{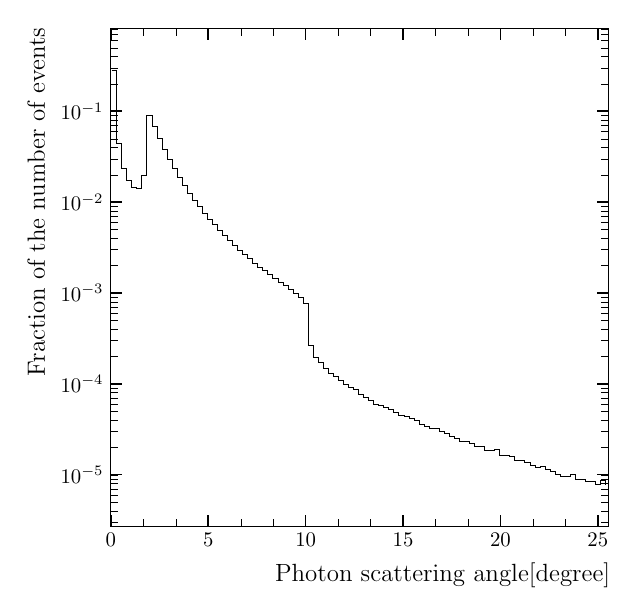} & \includegraphics[scale=0.8]{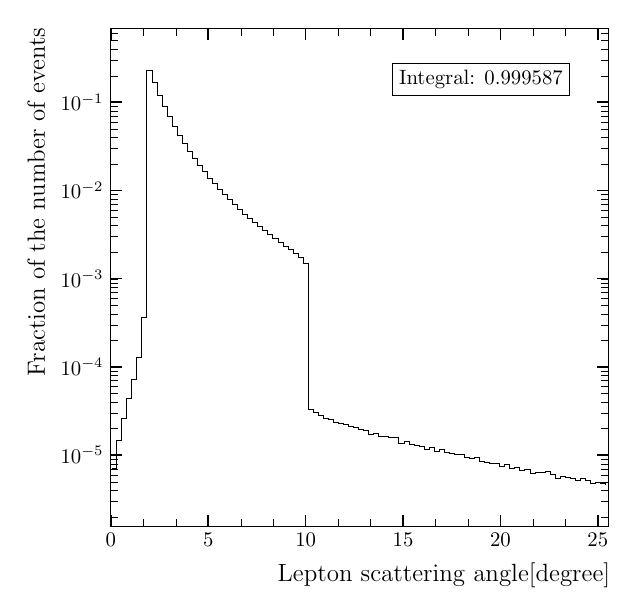}\\}; 
    \end{tikzpicture}
    \caption{Angular distributions of type a) on the left side and type b) on the right side for $\sqrt{s} = 240$ GeV.}
    \label{fig:3}
\end{figure}

\begin{figure}[h!]
\begin{tikzpicture}
    \matrix[matrix of nodes]{
        \includegraphics[scale=0.8]{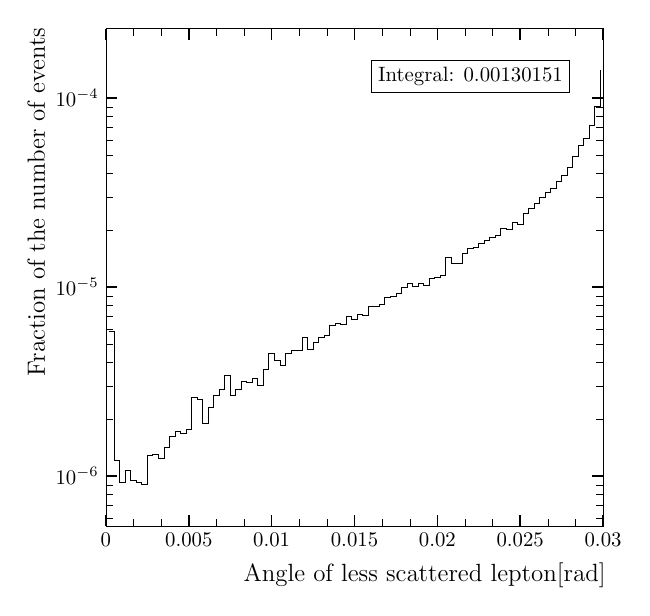} &
        \includegraphics[scale=0.8]{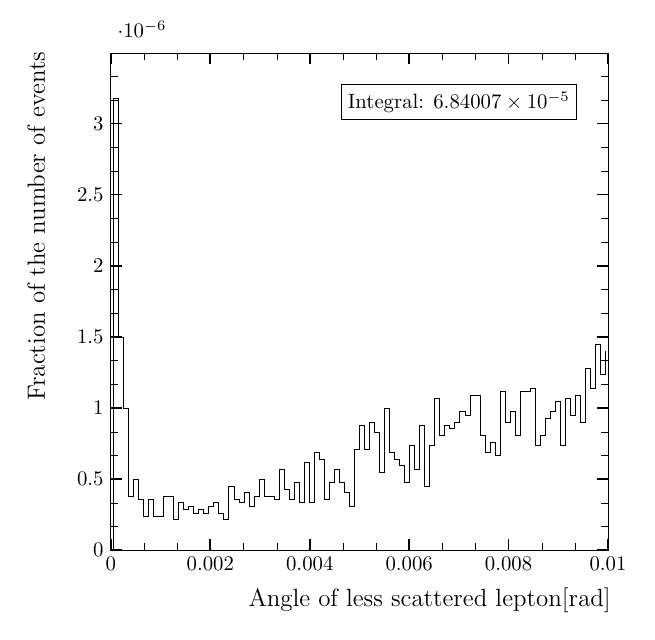} &
        \\};
    \end{tikzpicture}
    \caption{
    Angular distributions of type a) on the left side and type b) on the right side for $\sqrt{s} = 240$ GeV.}
    \label{fig:4}
\end{figure}

Fig. \ref{fig:3}--\ref{fig:4} show the same angular distributions for the 240~GeV collision energy.
The relative event yields are very similar to the case of $Z$ resonance, i.e., 1.3 permille below 30 mrad, and less than $10^{-4}$ below 10 mrad.

\vspace*{3mm}

\underline{18.9 mrad angular cut-off} \\

OPAL experiment at LEP has partially taken into account the effect of very low angle scattering
of electrons by generating
events with 18.9 mrad minimum angular cut-off \cite{OPAL:1999clt}, which is considerably lower than the experimental acceptance domain.
The contribution of scattering by smaller angles was estimated by extrapolation to be less than $2\cdot 10^{-5}$ and was neglected. But the simple extrapolation could underestimate
the neglected contribution because of the peak at extremely small angles which is seen
in Fig.~\ref{fig:4}.
Our calculations show
that the neglected contribution amounts to about $2.3 \cdot 10^{-4}$ at 91~GeV collision energy.
This is still within the total theoretical uncertainty of $5.4\cdot 10^{-4}$ assumed 
in \cite{OPAL:1999clt}\footnote{Taking into account this effect would have enlarged the resulting uncertainty.}.

\section{Conclusions}

In this way, we applied the Monte Carlo {\tt MCSANC} integrator and {\tt ReneSANCe} generator
for description of small-angle Bhabha scattering. We verified that the results
of the two programs are consistent with each other within statistical errors.
At the level of one-loop QED radiative corrections their results agree also
with the ones of the BHLUMI event generator~\cite{Jadach:1996is}. 
We took into account also the leading effects due to higher-order electroweak corrections 
and vacuum polarization. 
We examined SABS as a possible process to monitor 
the luminosity at future $e^+e^-$ experiments aiming at the $10^{-4}$
level of uncertainty.
Here we limited ourselves to considering only perturbative effects, whereas in a realistic 
situation other effects must be taken into account, for example, beamstrahlung
and the final size of the beams~\cite{Baier:2002rg,Kotkin:2003jz} must be taken into account.

The unique feature of the {\tt SANC} tools allows one to generate radiative 
Bhabha events with electron scattering angles down to zero.
This allowed us to take into account events in which one arm of the luminosity 
calorimeter is fired by an energetic ISR photon, while an electron is scattered 
by very small angle and escapes detection. 
Based on calculations of both {\tt MCSANC} integrator and {\tt ReneSANCe} generator we observe a contribution of 1.3-1.4 permille from events
with scattering angles less than 30 mrad, both at $Z$ pole and at 240~GeV.
This effect represent a significant bias given the high experimental precision 
expected at the future colliders. 
The bias can influence in particular the measurements of the total 
luminosity, the effective number of neutrino flavors $N_\nu$, etc.
To meet the expected precision of the future colliders 
($\sim 10^{-4}$), we recommend to generate events with angular cut-off somewhat less than 10~mrad, 
or to use generators capable to generate zero scattering angles.

In order to match the required uncertainty, we have to implement in our codes the complete two-loop
QED radiative corrections to Bhabha scattering and the leading and sub-leading contributions
enhanced by large logarithms.

\addcontentsline{toc}{section}{\refname}
\printbibliography

\end{document}